\documentstyle[prl,aps]{revtex}

\makeatletter
\def\lsim{\mathrel{\mathpalette\@versim<}}
\def\gsim{\mathrel{\mathpalette\@versim>}}
\def\@versim#1#2{\vcenter{\offinterlineskip
        \ialign{$\m@th#1\hfil##\hfil$\crcr#2\crcr\sim\crcr } }}
\makeatother

\def\pl{P}
\def\R{{\rm R}}
\def\L{{\rm L}}
\def\pl{{\rm Pl}}
\def\m1{{-1}}

\def\VEV#1{\left\langle#1\right\rangle}

\newcommand{\nn}{\nonumber\\}

\def\gam#1{\,{}^{#1}\kern-1pt\Gamma}

\def\tpmatrix#1{\left(\begin{array}{@{\,}c@{\,\ }c@{\,}}
#1\end{array}\right)}
\begin{document}
\draft
\twocolumn[\hsize\textwidth\columnwidth\hsize\csname 
@twocolumnfalse\endcsname 

\title{Implications of Bi-Large Neutrino Mixing on GUTs} 
\author{Masako Bando and Taichiro Kugo$^1$}
\address{
  Physics Division, Aichi University, Aichi 470-0296, Japan\\ 
  $^1$Yukawa Institute, Kyoto University, Kyoto 606-8502, Japan}
\date{\today}

~{}\vskip-2ex
YITP-03-49 \hfill  
{\tt hep-ph/0308258}

\maketitle

\begin{abstract} 
Under the assumptions that 1) the quark/lepton mass matrices take
Froggatt-Nielsen's factorized power form $\lambda^{\psi_i+\psi_j}$
with anomalous $U(1)$ charges $\psi_i$, and 2) the $U(1)$ charges
$\psi_i$ respect the $SU(5)$ GUT structure, we show that the quark
mass data and the mass-squared difference ratio of solar and
atmospheric neutrinos, as inputs, necessarily imply that both the 1-2
and 2-3 mixings in the MNS matrix $U_{\rm MNS}$ are large. This
analysis also gives a prediction that $U_{e3}\equiv(U_{\rm MNS})_{13}$
is of order $\lambda\,\sim\, (0.1 - 0.5)$. We also add an argument
that $E_6$ GUT is favored. \end{abstract} 

]

Existence of a certain grand unified theory (GUT) beyond the standard 
model is guaranteed by i) the anomaly cancellation between quarks and 
leptons and ii) the unification of the gauge coupling constants at 
energy scale around $\mu\,\sim\,10^{16}$\,GeV. The strongest candidate 
for the unified gauge group is $E_6$, which is not only 
suggested by string theory but also unique in the property that it is 
the maximal safe simple group allowing 
complex representations in the $E$-series; $E_3=SU(3)\times SU(2)$, 
$E_4=SU(5)$, $E_5=SO(10)$, $E_6, \ E_7,\ E_8$.\cite{ref:Georgi}

On the other hand, the present neutrino data
\cite{SuperKam:1998,SuperKam:2002,SNO,Kamland} 
show the following particular facts:

1. Bi-large mixing;
\begin{equation}
\sin^22\theta_{12} \  \sim\ 0.8, \qquad 
\sin^22\theta_{23} \  \sim\ 1.
\end{equation}

2. Mass-squared difference ratio of solar ($\odot$) to
 atmospheric ($\oplus$) neutrinos;
\begin{equation}
{\triangle m_\odot^2\over\triangle m_{\oplus}^2} 
\ \sim\ {7\times10^{-5}\,{\rm eV}^2\over 
2\times10^{-3}\,{\rm eV}^2} 
\ \sim\ \lambda^2,
\label{eq:dmRatio}
\end{equation}
where $\lambda$ defined below is a quantity of magnitude 
$\lambda\sim0.22$.

These show a sharp contrast to the quark sector, in which 
the mixings are very small and the mass spectrum is hierarchical. 
The mutual relations of the masses and mixing angles between 
quarks and leptons/neutrinos will be great clues for the GUTs. 

The purpose of this paper is to analyze the implications of these neutrino 
data on the possible GUTs.\cite{ref:NOON2003} We analyze these data 
first assuming, as an working hypothesis, a supersymmetric SU(5) GUT and 
the Froggatt-Nielsen mechanism\cite{Froggatt:1978nt} for generating 
hierarchical quark/lepton masses. 
The latter mechanism utilizes a (usually anomalous) $U(1)_X$ 
charge to generate effective Yukawa couplings via higher 
dimensional interaction terms in the superpotential of the form 
\begin{equation}
y\Psi_i\Psi_jH\left({\Theta\over M_\pl}\right)^{\psi_i+\psi_j+h},
\label{eq:FNcoupling}
\end{equation}
where the `pre-Yukawa' coupling constants $y$ can in principle 
depend on the generation 
label $i,j$ but are here assumed to be all of order 1 and so are denoted by
$y$ collectively. $\Theta$ is the Froggatt-Nielsen field carrying the 
$U(1)_X$ charge $-1$ and the $U(1)_X$ charges of the Higgs chiral 
superfield $H$ and matter chiral superfields $\Psi_i$ ($i=1,2,3$) are 
denoted by the corresponding lower-case letters: 
\begin{equation}
X(\Theta)=-1, \qquad X(H)=h, \qquad X(\Psi_i)=\psi_i\,(\,\geq0).
\end{equation}
After the  Froggatt-Nielsen field $\Theta$ develops a vacuum expectation 
value (VEV) $\VEV{\Theta}$, which is assumed to be smaller than 
the Planck scale by a factor of Cabibbo angle $\theta_{\rm C}$
\begin{equation}
{\VEV{\Theta}\over M_\pl} \equiv\lambda\ \sim\ 0.22\simeq \sin\theta_{\rm C},
\label{eq:VEV}
\end{equation}
the effective Yukawa couplings induced from Eq.~(\ref{eq:FNcoupling})
are given by
\begin{equation}
y_{ij}^{\rm eff}= y \times\lambda^{\psi_i+\psi_j+h}= {\cal O}(1)\times\lambda^{\psi_i+\psi_j+h}.
\end{equation}

That is, 
suppressing the ${\cal O}(1)$ coefficients henceforth, and 
distingushing the right-handed and left-handed matter superfields 
$\Psi_i^\R$ and $\Psi_i^\L$, 
the mass matrix $M$ takes the form
\begin{equation}
M = yv\lambda^h \times\bordermatrix{
 & &  \stackrel{j}{\vee} & \cr
 & &  & \cr
i{>} & & \lambda^{\psi^\R_i+\psi^\L_j}  & \cr
 & &  & \cr}
\end{equation}
with $\VEV{H}=v$ 
and $\psi^\R_i$ and $\psi^\L_j$ denoting the $U(1)_X$ charges of 
$\Psi_i^\R$ and $\Psi_i^\L$, respectively.
Thus, in this Froggatt-Nielsen mechanism, the hierarchical mass 
structure can be explained by the difference of the $U(1)_X$ charges $\psi 
^{\R,\L}_i$ of the matter fields. Note that this type of `factorized' 
mass matrix can be diagonalized as 
\begin{equation}
VMU^\dagger= M^{\rm diag.}
\end{equation} 
by the unitary matrices $U$ and $V$ taking also similar power forms:
\begin{equation}
U \, \sim\, 
\Bigl(\ 
\lambda^{\left|\psi^\L_i-\psi^\L_j\right|} 
\ \Bigr),\quad 
V \, \sim\, 
\Bigl(\ 
\lambda^{\left|\psi^\R_i-\psi^\R_j\right|} 
\ \Bigr).
\end{equation}

We assume $SU(5)$ structure for the quark and lepton matter contents 
and the $U(1)_X$ charge assignment for them. 
Then, the higher dimensional Yukawa couplings 
responsible for the up-quark sector masses, which are invariant under 
$SU(5)$ and $U(1)_X$, are given by:
\begin{equation}
y_u\Psi_i({\bf10})\Psi_j({\bf10})H_u({\bf5})
\left({\Theta\over M_\pl}\right)^{\psi_i({\bf10})+\psi_j({\bf10})+h_u}. 
\end{equation}
After the VEV (\ref{eq:VEV}) is developed,
these yield  the effective Yukawa coupling constants
\begin{eqnarray}
{y_u}_{ij}^{\rm eff}=y_u \times\lambda^{\psi_i({\bf10})+\psi_j({\bf10})+h_u}. 
\end{eqnarray}
In order for these to reproduce the observed up-type quark mass 
hierarchy structure 
\begin{equation}
m_t : m_c : m_u 
\mathrel{\mathop{\kern0pt =}\limits_{\hbox{\scriptsize exp.}}}\ 
1 : \lambda^4 : \lambda^7\ ,
\end{equation}
we are led to choose the following values for the $U(1)_X$ charges of 
three generation $\Psi_i({\bf10})$ fermions 
taking $h_u=0$ for simplicity:\cite{ref:BK}
\begin{equation}
(\,\psi_1({\bf10}),\ \psi_2({\bf10}),\ \psi_3({\bf10})\,)
=(\,3,\ 2,\ 0\,)
\label{eq:value10}
\end{equation}

Next we consider the mass matrices of down-type quarks and charged 
leptons which come from the couplings
\begin{eqnarray}
&&y_d\Psi_i({\bf10})\Psi_j({\bf5}^*)H_d({\bf5}^*)
\left({\Theta\over M_\pl}\right)^{\psi_i({\bf10})+\psi_j({\bf5}^*)+h_d}
\nn && \hspace{2em} 
 \rightarrow\ 
{y_d}_{ij}^{\rm eff}= y_d \times\lambda^{\psi_i({\bf10})+\psi_j({\bf5}^*)+h_d}. 
\end{eqnarray}
Note that this yields the transposed relation between the 
 down-type quark mass matrix $M_d$ and the charged lepton one $M_l$:
$
{M_d}^T  \sim M_l 
$. 
This is because the $\Psi_i({\bf5}^*)$ multiplets contain the 
right-handed component $d^c$ for the down-type quarks while 
the left-handed component $l$ for the charged leptons.
Therefore the unitary matrices for diagonalizing those mass matrices, 
satisfy the relations
\begin{equation}
\left\{
\begin{array}{@{\,}l}
V_dM_dU_d^\dagger= M_d^{\rm diag.} \\[.5ex]
V_lM_lU_l^\dagger= M_l^{\rm diag.}
\end{array}\right. \quad \rightarrow\quad 
\left\{
\begin{array}{@{\,}l}
V_l=U_d^* \\[.5ex]
V_d=U_l^* 
\end{array}\right. ,
\end{equation}
so that we have 
$U_d^* (M_l\sim M_d^T) U_l^\dagger= \hbox{diag.}$ 
with
\begin{equation}
U_d \sim\Bigl( 
\lambda^{\left|\psi_i({\bf10})-\psi_j({\bf10})\right|}\Bigr), \ \  
U_l  \sim\Bigl( 
\lambda^{\left|\psi_i({\bf5}^*)-\psi_j({\bf5}^*)\right|}\Bigr).
\end{equation}
That is, the mass matrix takes the form
\begin{eqnarray}
&&M_d^T \sim M_l  \sim 
y_dv\lambda^{h_d} \times\Bigl(\ 
  \lambda^{\psi_i({\bf10})+\psi_j({\bf5}^*)} 
\ \Bigr)\,.
\label{eq:dlmass}
\end{eqnarray}
In order for this $M_d$ to reproduce the mass ratio of the top and bottom 
quarks
\begin{equation}
{m_b\over m_t} \ 
 \mathrel{\mathop{\kern0pt \sim}\limits_{\hbox{\scriptsize exp.}}}\ 
\lambda^{2 - 3}
\end{equation}
we take $\psi_3({\bf5}^*)=2-h_d$.
Further, to reproduce 
the down-type quark mass hierarchy
\begin{equation}
m_b : m_s : m_d 
\mathrel{\mathop{\kern0pt =}\limits_{\hbox{\scriptsize exp.}}}\ 
1 : \lambda^2 : \lambda^4\ ,
\end{equation}
we take $\psi_2({\bf5}^*)=\psi_1({\bf5}^*)-1=\psi_3({\bf5}^*)$, so that 
\begin{equation}
(\,\psi_1({\bf5}^*),\ \psi_2({\bf5}^*),\ \psi_3({\bf5}^*)\,)
=(\,3-h_d,\ 2-h_d,\ 2-h_d\,), 
\label{eq:value5}
\end{equation}
and the mass matrix (\ref{eq:dlmass}) now reduces to 
\begin{equation}
M_d^T \sim M_l \  \sim\ 
y_dv\lambda^2 \times 
\pmatrix{
\lambda^4 & \lambda^{3} & \lambda^{3} \cr
\lambda^{3} & \lambda^{2} & \lambda^{2} \cr
\lambda &  1  & 1 \cr} 
\end{equation}
This form of mass matrix is called lopsided.


Mixing matrices in the quark sector and lepton sector are called 
Cabibbo-Kobayashi-Maskawa (CKM) 
and Maki-Nakagawa-Sakata (MNS)\cite{ref:MNS} 
matrices, respectively, and they are defined by 
\begin{equation}
U_{\rm CKM} = U_uU_d^\dagger\,,\qquad 
U_{\rm MNS} = U_lU_\nu^\dagger\,.
\end{equation} 
In our case both $U_u$ and $U_d$ takes the form 
$U_u\,\sim\,U_d\,\sim\,
( \lambda^{\left|\psi_i({\bf10})-\psi_j({\bf10})\right|})$, 
so that the CKM matrix, generally, also has the same form
\begin{equation}
U_{\rm CKM}
\, \sim\, 
\Bigl( \lambda^{\left|\psi_i({\bf10})-\psi_j({\bf10})\right|}\Bigr)
\, \sim\, 
\pmatrix{
1 & \lambda& \lambda^3 \cr
\lambda& 1 & \lambda^2 \cr
\lambda^3 & \lambda^2 & 1 \cr},
\end{equation}
agreeing perfectly the experimatal data.  
For the charged lepton sector we have 
\begin{equation}
U_l \ \sim\ 
\Bigl( \lambda^{\left|\psi_i({\bf5}^*)-\psi_j({\bf5}^*)\right|}\Bigr)
\ \sim\ 
\pmatrix{
1 & \lambda& \lambda\cr
\lambda& 1 & 1  \cr
\lambda& 1 & 1 \cr}.
\end{equation}
If the mixing matrix $U_\nu$ in neutrino sector is $\sim1$, 
this beautifully explains the observed 
large 2-3 neutrino mixing!  However, this alone fails in explaining 
the large 1-2 mixing. We thus have to discuss the 
neutrino mixing matrix $U_\nu$ now. 

%


Generally in GUTs, there appear some right-handed neutrinos 
$\Psi_I({\bf1})=\nu_{\R I}$ ($I=1,\cdots,n$); 
for instance, $n=3$ in $SO(10)$ and $n=6$ in 
$E_6$.\cite{ref:BK} They will generally get superheavy 
Majorana masses denoted by an $n\times n$ 
mass matrix $(M_\R)_{IJ}$, 
and also possesses the Dirac masses (R-L transition mass terms) 
\begin{equation}
\bigl(M_{\rm D}^T\bigr)_{iI} \  \sim\ 
y_\nu v\lambda^{h_u} \times\Bigl(
\lambda^{\psi_i({\bf5}^*)+\psi_I^\R} \Bigr)
\end{equation}
induced from 
\begin{eqnarray}
&&y_\nu\Psi_i({\bf5}^*)\Psi_I({\bf1})H_u({\bf5})
\left({\Theta\over M_\pl}\right)^{\psi_i({\bf5}^*)+\psi_I^\R+h_u}.
\end{eqnarray}
Here $\psi_I^\R$ denotes the $U(1)_X$ charges of the right-handed 
neutrinos $\Psi_I({\bf1})$. 

The Majorana mass matrix $M_\nu$ of (left-handed) neutrino is induced 
from these masses $M_\R$ and $M_{\rm D}$ by 
the see-saw mechanism\cite{ref:seesaw} 
and evaluated as
\begin{eqnarray}
\bigl(M_{\nu}\bigr)_{ij}
&\sim& \bigl(M_{\rm D}^T\bigr)_{iI}
\bigl(M_\R^{-1}\bigr)_{IJ}
\bigl(M_{\rm D}\bigr)_{Jj} \nn
 &\sim&
\lambda^{\psi_i({\bf5}^*)}\left(\lambda^{\psi_I^\R}
\bigl(M_\R^{-1}\bigr)_{IJ}
\lambda^{\psi_J^\R}\right)
\lambda^{\psi_j({\bf5}^*)}\nn
&\propto&
\lambda^{\psi_i({\bf5}^*)+\psi_j({\bf5}^*)}\ .
\label{eq:seesaw}
\end{eqnarray}
Note that the dependence on the 
$U(1)_X$ charges $\psi_I^\R$ of the right-handed 
neutrinos has completely dropped out. (We should however take it into
account that this occurs only for a generic case and may be broken in 
particular cases in which $\bigl(M_{\rm D}^T\bigr)_{iI}$ brings about 
correlation between the left-handed neutrino index $i$ and right-handed 
one $I$.\cite{ref:BK}) \  Plaguing the values (\ref{eq:value5}) for $\psi 
_i({\bf5}^*)$, we thus have
\begin{equation}
M_\nu\  \propto\  
\pmatrix{
\lambda^2 & \lambda& \lambda\cr
\lambda& 1 & 1 \cr
\lambda& 1 & 1 \cr}.
\label{eq:Ours}
\end{equation}
This neutrino mass matrix happens to take the same form as one of the 
models that have been proposed by Ling and Ramond\cite{ref:Ramond} 
and Babu, Gogoladze and Wang\cite{Babu:2002tx}
This form is very interesting. 

First, this matrix implies the large 2-3 mixing in the 
diagonalization matrix $U_\nu$. The 2-3 mixing is also large in the 
charged lepton mixing matrix $U_l$ as we have seen above, and so is it 
generally in the MNS matrix $U_{\rm MNS}=U_lU_\nu^\dagger$ unless a 
cancellation occurs between $U_l$ and $U_\nu$. 

Second, it is natural to assume that three neutrino masses are
not accidentally degenerate . Then, the mass squared difference ratio 
(\ref{eq:dmRatio}) of the solar and atmospheric neutrinos 
implies the mass ratio of the second and third neutrinos:
$m_{\nu2}/m_{\nu3} \sim\lambda$.
In order for the $M_\nu$ to reproduce this mass ratio, 
the determinant of the $2\times2$ bottom-right submatrix 
of this $M_\nu$ should not be naturally-expected order 1, but 
should be $O(\lambda)$; that is, that submatrix should be diagonalized 
by an $2\times2$ unitary matrix $u_\nu$ as
\begin{equation}
u^*_\nu\tpmatrix{
 1 & 1 \\
 1 & 1 } u^\dagger_\nu\,\sim\, 
\pmatrix{  \lambda& 0 \cr
 0 & 1 \cr}.
\end{equation}
If this is the case, 
the mass matrix $M_\nu$ takes the following form after the 
diagonalization of this $2\times2$ bottom-right submatrix:
\begin{equation}
\pmatrix{
1 & 0 \cr 
0 & u_\nu^* \cr}
M_\nu 
\pmatrix{
1 & 0 \cr 
0 & u_\nu^\dagger\cr} 
\, \sim\, 
\pmatrix{
\lambda^2 & \lambda& \lambda\cr
\lambda& \lambda& 0 \cr
\lambda& 0 & 1 \cr}.
\end{equation}
If we note the $2\times2$ top-left submatrix of this matrix
\begin{equation}
\pmatrix{
\lambda^2 & \lambda\cr
\lambda& \lambda\cr},
\end{equation}
we see that this also gives the large mixing in the 1-2 sector so that 
it explains the {\em bi-large mixing}. 

Therefore, the experimental fact 
\begin{equation}
{\triangle m_\odot^2\over\triangle m_{\oplus}^2} \ \sim\ \lambda^2
\qquad \Leftrightarrow\qquad 
{m_{\nu2}\over m_{\nu3}}\ \sim\ \lambda 
\end{equation}
necessarily implies the {\rm bi-large mixing}!

We note that a very similar neutrino mass matrix $M_\nu$ to ours
(\ref{eq:Ours}) was also proposed by Maekawa:\cite{ref:maekawa}
\begin{equation}
M_\nu\  \propto\  
\pmatrix{
\lambda^2 & \lambda^{1.5} & \lambda^1 \cr
\lambda^{1.5} & \lambda^1 & \lambda^{0.5} \cr
\lambda^1 & \lambda^{0.5} & 1 \cr}.
\label{eq:Maekawa}
\end{equation}


We should note that there is one more prediction in our framework; 
that is, it predicts a rather `large' value for the element 
$U_{e3}\equiv(U_{\rm MNS})_{13}$:
\begin{equation}
U_{e3} \ \sim\ O(\lambda^1) \quad  \sim\quad  
\bigl(\ {0.1} - {0.5}\ \bigr)\,.
\end{equation}
This is seen as follows.
First, for $U_l$, we have
\begin{eqnarray}
(U_l)_{11} \ &\sim&\ O(1), \nn
(U_l)_{12} \ \hbox{and} \ (U_l)_{13} \ &\sim&\  
\lambda^{\psi_1({\bf5}^*)-\psi_{{2\atop3}}({\bf5}^*)} = \lambda^1,
\end{eqnarray}
which have resulted from down-type quark masses and an 
$SU(5)$ relation. Second, for $U_\nu$, we have 
\begin{eqnarray}
&&(U_\nu)_{31} \ \sim\ \lambda^{\psi_1({\bf5}^*)-\psi_3({\bf5}^*)} = \lambda^1, \nn
&&(U_\nu)_{32} \ \hbox{and} \ (U_\nu)_{33} \ \sim\ O(1).
\end{eqnarray}
These clearly give rise to 
$U_{e3}\equiv(U_{\rm MNS})_{13}=(U_lU_\nu^\dagger)_{13}\sim O(\lambda)$. 
Although the bigger side of this prediction is already excluded 
experimentally, this prediction gives a crucial test for the idea of 
Froggatt-Nielsen mechanism.

Summarizing the points up to here, we have shown:

1. If we assume Froggatt-Nielsen's factorized form for the quark/lepton 
mass matrices and the $SU(5)$ structure for the $U(1)_X$ charges, an 
input of up- and down-type quark masses necessarily implies that 
the 2-3 mixing is large in the MNS matrix $U_{\rm MNS}$.

2. If we further add the data $\sqrt{\triangle m_\odot^2/\triangle m_{\oplus}^2} \ \sim\ \lambda$,
then, it implies that the 1-2 mixing in $U_{\rm MNS}$ is also large, 
so leading to bi-large mixing.

3. The measurement of $U_{e3}$ will {\em confirm} or {\em kill} the 
basic idea of {\em Froggatt-Nielsen mechanism} for explaining the 
hierarchical mass structures of quarks and leptons. 

Let us now look back our analysis and examine the possible implication 
of the neutrino data on the GUTs. 

Recall first that the mixing unitary matrices are determined solely by 
the $U(1)$ charges of the left-handed components, and that the 
left-handed components fall into $SU(2)$ doublets under the standard 
gauge symmetry. Therefore, under the assumption of the generic 
factorized power form for the mass matrices, the standard gauge symmetry
alone predicts 
\begin{equation}
U_u \sim U_d \sim\Bigl( \lambda^{|Q_i-Q_j|} \Bigr), \ \ 
U_\nu\sim U_l \sim\Bigl( \lambda^{|L_i-L_j|} \Bigr),
\end{equation}
and hence also, for CKM and MNS matrices, 
\begin{eqnarray}
U_{\rm CKM} &=& U_uU_d^\dagger\,\sim\, \Bigl( \lambda^{|Q_i-Q_j|} \Bigr),  \nn
U_{\rm MNS} &=& U_lU_\nu^\dagger\,\sim\, \Bigl( \lambda^{|L_i-L_j|} \Bigr)\,.
\end{eqnarray}
Here $Q_i$ and $L_i$ are the U(1) charges of the quark and lepton 
doublets, respectively.
If we had the Pati-Salam symmetry $SU(4)_{\rm PS}$, instead of 
the $SU(5)$ symmetry as assumed in the above, the $U(1)$ charges 
of quark and lepton doublets must be the same, $Q_i=L_i$, 
which leads to an incorrect prediction that the
CKM and MNS mixing matrices should have the same structure, 
$U_{\rm CKM} \sim U_{\rm MNS}$. 

This apparently seems to exclude GUT gauge groups like SO(10) and $E_6$ 
larger than SU(5), since they necessarily contain the $SU(4)_{\rm PS}$ and 
the gauge multiplet members all carry a common $U(1)_X$ charge. 
Actually, any $SO(10)$ models in which three generations of matters 
come from three {\bf 16} representations are excluded, since there the 
leptons are the fourth colored `quarks' of $SU(4)_{\rm PS}$ 
and so the relations $Q_i=L_i$ must hold.
The group $E_6$, however, has two intrinsic mechanisms 
giving the ways out of this difficulty, and hence   
the $E_6$ models in which three generations come from three 
{\bf 27} representations are indeed allowed. 

The first reason is because {\bf27} is decomposed into the $SU(5)$ 
multiplets 
\begin{equation}
{\bf27}= 
\underbrace{({\bf10}+{\bf5}^*+{\bf1})}_{SO(10)\,{\bf16}} +
\underbrace{({\bf5}+{\bf5}^*)}_{SO(10)\,{\bf10}} +
\underbrace{{\bf1}}_{SO(10)\,{\bf1}} 
\end{equation}
The point here is that there appear two $SU(5)$ ${\bf5}^*$ representatioons 
in each {\bf27}, and six in all for three generation case. 
The $SU(4)_{\rm PS}$ partner of the quark doublet contained in 
{\bf10} in $SO(10)\,{\bf16}$ is the lepton doublet in 
the ${\bf5}^*$ in $SO(10)\,{\bf16}$, but not that in the 
the ${\bf5}^*$ in $SO(10)\,{\bf10}$. However, the light lepton doublets 
may generally come from any three (linear combinations) out of these six 
$SU(5)$ ${\bf5}^*$, and, so the relation $Q_i=L_i$ can easily be 
avoided.\cite{ref:BK,BandoMaekawa}

Second, it is also important that three {\bf27} contain six 
$SU(5)$ singlets ${\bf1}$ which play the roll of right-handed neutrinos
and form $6\times3$ Dirac mass matrix with the three left-handed neutrinos 
in the three light lepton doublets. However, depending on the $SO(10)$ 
properties of those three lepton doublets, there may appear zeros in the
matrix elements of the $6\times3$ Dirac mass matrix. 
This is because, for instance, $SU(5)$
singlet in $SO(10)$ {\bf16}, denoted 
$({\bf1}, {\bf16})$ for brevity, 
can form Dirac masses with the left-handed neutrinos in $({\bf5}^*, 
{\bf16})$ but not those in $({\bf5}^*, {\bf10})$, when the Higgs is 
assumed to be $({\bf5},{\bf10})$. 
And $6\times6$ Majorana mass
matrix for the right-handed neutrinos can have additional structure 
other than the Froggatt-Nielsen's $U(1)$ power structure, since 
$({\bf1},{\bf16})$ and $({\bf1},{\bf1})$ have different $SO(10)$ quantum
numbers and their Majorana masses may come from different Higgs scalars.
These two facts invalidate the above discusion in Eq.~(\ref{eq:seesaw}) 
proving the factorization of the light neutrino mass matrix $M_\nu$. So 
$U_\nu\sim\Bigl( \lambda^{|L_i-L_j|} 
\Bigr)$ no longer remains true and $U_{\rm CKM} \not\sim U_{\rm MNS}$ even
when $Q_i=L_i$.




We would like to thank N.~Maekawa 
for valuable discussions. We were also inspired by stimulating 
discussions during the Summer 
Institute 2002 and 2003 held at Fuji-Yoshida.


\begin{thebibliography}{99}

\bibitem{ref:Georgi}
{\em Lie Algebras in Particle Physics},  (Westview Press, 1999)


\bibitem{SuperKam:1998}
Y.~Fukuda {\it et al.}  [Super-Kamiokande Collaboration],
Phys.\ Lett.\ B {\bf 433}, 9 (1998)
[arXiv:hep-ex/9803006]; 
Phys.\ Lett.\ B {\bf 436}, 33 (1998)
[arXiv:hep-ex/9805006].

\bibitem{SuperKam:2002}
S.~Fukuda {\it et al.}  [Super-Kamiokande Collaboration],
Phys.\ Lett.\ B {\bf 539}, 179 (2002)
[arXiv:hep-ex/0205075]; 
Phys.\ Rev.\ Lett.\  {\bf 86}, 5651 (2001)
[arXiv:hep-ex/0103032];
Phys.\ Rev.\ Lett.\  {\bf 86}, 5656 (2001)
[arXiv:hep-ex/0103033].

\bibitem{SNO}
Q.~R.~Ahmad {\it et al.}  [SNO Collaboration],
Observatory,''
Phys.\ Rev.\ Lett.\  {\bf 89}, 011301 (2002)
[arXiv:nucl-ex/0204008]; 
Phys.\ Rev.\ Lett.\  {\bf 89}, 011302 (2002)
[arXiv:nucl-ex/0204009].

\bibitem{Kamland}
K.~Eguchi {\it et al.}  [KamLAND Collaboration],
Phys.\ Rev.\ Lett.\  {\bf 90}, 021802 (2003)
[arXiv:hep-ex/0212021].

\bibitem{ref:NOON2003}
A part of the results of this work was 
already reported by T.~Kugo in NOON 2003.

\bibitem{Froggatt:1978nt}
C.~D.~Froggatt and H.~B.~Nielsen,
Nucl.\ Phys.\ B {\bf 147}, 277 (1979).

\bibitem{ref:BK}
M.~Bando and T.~Kugo,
Prog.\ Theor.\ Phys.\  {\bf 101}, 1313 (1999) 
[arXiv:hep-ph/9902204]. \\
M.~Bando, T.~Kugo and K.~Yoshioka,
Prog.\ Theor.\ Phys.\  {\bf 104}, 211 (2000)
[arXiv:hep-ph/0003220].

\bibitem{ref:MNS}
	Z.~Maki, M.~Nakagawa and S.~Sakata, 
Prog.\ Theor.\ Phys.\  {\bf 28}, 870  (1962).

\bibitem{ref:seesaw}
  T.~Yanagida, in {\sl Proceedings of the Workshop on Unified Theory
    and Baryon Number of the Universe}, eds.\ O.~Sawada and
  A.~Sugamoto (KEK, 1979); M.~Gell-Mann, P.~Ramond and R.~Slansky, in
  {\sl Supergravity}, eds.\ P.~van Nieuwenhuizen and D.Z.~Freedman
  (North Holland, Amsterdam, 1979). 

\bibitem{ref:Ramond}
F.~S.~Ling and P.~Ramond,
Phys.\ Lett.\ B {\bf 543}, 29 (2002) 
[arXiv:hep-ph/0206004].

\bibitem{Babu:2002tx}
K.~S.~Babu, I.~Gogoladze and K.~Wang,
Nucl.\ Phys.\ B {\bf 660}, 322 (2003)
[arXiv:hep-ph/0212245], and references therein.

\bibitem{ref:maekawa}   
N.~Maekawa, 
Prog.\ Theor.\ Phys.\  {\bf 106}, 401 (2001)
                       [arXiv:hep-ph/0104200]; 
Prog.\ Theor.\ Phys.\  {\bf 107}, 597 (2002)
                       [arXiv:hep-ph/0111205]. 
\bibitem{BandoMaekawa}   
M.~Bando and N.~Maekawa,
Prog.\ Theor.\ Phys.\  {\bf 106}, 1255 (2001) 
[arXiv:hep-ph/0109018].
%

\end{thebibliography}
\end{document}